\documentclass{aa}

\usepackage{graphicx}
\usepackage{txfonts}
\usepackage{natbib}
\usepackage{color}
\usepackage{fancyvrb}

\title{On the visibility of stellar oscillations}

\author{J. Schou}

\date{\today}

\institute{
Max-Planck-Institut f\"ur Sonnensystemforschung, Justus-von-Liebig-Weg
3, 37077 G\"ottingen, Germany\\
\email{schou@mps.mpg.de}
}

\begin{document} 

 
\abstract
{
Recently our ability to study stars using asteroseismic techniques has increased dramatically, largely through the use of space based photometric observations. Work has also been done using ground based spectroscopic observations and more is expected in the near future from the SONG network.
Unfortunately, the intensity observations have an inferior signal to noise ratio and details of the observations do not agree with theory, while the data analysis used in the spectroscopic method has often been based on overly simple models of the spectra.
}
{
The aim is to improve the reliability of measurements of the
parameters of stellar oscillations using spectroscopic observations
and to enable the optimal use of the observations.
}
{
While previous investigations have used 1D models,
I use realistic magnetohydrodynamic simulations, combined
with radiative transfer calculations, to estimate the
effects of the oscillations on the spectra. I then calculate the
visibility of the oscillation modes for a variety of stellar
parameters and using various fitting methods.
In addition to the methods used in previous investigations, I use a
Singular Value Decomposition technique, which allows one to determine the
information content available from the spectral perturbations
and how that can be expressed most compactly.
Finally I describe how the time series obtained may be analyzed.
}
{
It is shown that it is important to model the visibilities
carefully and that the results deviate substantially from models
previously used, especially in the presence of rotation.
Detailed spectral modeling may be exploited to measure the properties of
a larger number of modes than possible using the
commonly used cross-correlation method.
With moderate rotation,
there is as much information in the line shape changes as in the Doppler
shift and an outline of how to extract this is given.
}
{}

\keywords{Asteroseismology - Stars: oscillations - Techniques: spectroscopic - Line: profiles}

\maketitle
%

\section{Introduction}
Most asteroseismic results have so far been obtained from photometric
observations using a variety of space-borne instruments,
such as WIRE \citep{2006MNRAS.371..935F}, MOST \citep{2003PASP..115.1023W},
COROT \citep{2009A&A...506..411A} and Kepler \citep{2010Sci...327..977B}.
This will likely also be the case in the future with the launch
of TESS \citep{2014SPIE.9143E..20R} and PLATO \citep{2014ExA....38..249R}.
The main advantage of photometric observations is that it is possible to observe
multiple stars simultaneously, as evidenced by the large number of main sequence stars for which
results have been obtained \citep{2013ApJ...765L..41S,2014ApJS..210....1C,2017ApJ...835..172L,2017ApJ...850..110L}.
However, Doppler velocity based
observations have a superior signal to noise ratio.
Unfortunately, it is difficult to observe more than one
star at a time with a spectrograph based
instrument and it is difficult to obtain the required
observing time (typically weeks to months at a high cadence
and high duty cycle on a large telescope) and therefore
only a few stars have been observed this way,
as reviewed by \cite{2014aste.book...60B}.
With the recent commissioning of the first SONG telescope \citep{2009CoAst.158..345G,2017ApJ...836..142G},
and the planned expansion of the SONG network,
more stars will be observed in the future.
Given this very limited resource and that the velocity observations are
significantly different from photometric observations, it is essential that the
maximum information is extracted from the observed stellar spectra and
that the resulting power spectra are accurately modeled.

When extracting the oscillation signal from the observed spectra,
Doppler shift algorithms developed for radial velocity
planet searches \citep[e.g.][]{2001A&A...374..733B}
are often used and it is thus implicitly
assumed that the oscillation signal is well represented by a Doppler
shift.
For fitting power spectra it is generally assumed that
the mode visibilities, as a function of the azimuthal order $m$,
follow the description of \cite{2003ApJ...589.1009G},
where it was assumed that the sensitivity only depends on the
center-to-limb distance on the stellar disk.
Below I will show that, in the presence of modest stellar
rotation, the Doppler shift approach discards a significant
amount of information and that the use of the visibility expression of
\cite{2003ApJ...589.1009G} can lead to
significant systematic errors.
The dependence of the visibilities on the spherical harmonic degree $l$
is typically kept free or prescribed to follow a relationship based on simplified models, even though it is known
that the observed visibilities for the Sun do not agree with the
theoretically expected values for broad band photometry
\citep{2014ApJ...782....2L}.

A few results, similar to those presented in the present paper, but using a different set of simulations were shown in
\cite{2014IAUS..301..481S}.
The present paper expands substantially on this by considering
stars of different spectral type, the effects of magnetic fields and a variety of analysis methods.

Models of the effects of oscillations
on spectra and how the information can be extracted have been studied
in the past, see for example \cite{1998ApJS..117..563B}, \cite{2003A&A...398..687B} and particularly \cite{2006A&A...455..227Z} where several physical effects
were included.
However, these models
have used simplified line models, rather than the results of three
dimensional magnetohydrodynamic (MHD) models, such as those used here
and the effects of magnetic fields on the spectra were not considered.
The use of a Singular Value Decomposition (SVD) based analysis was also not
discussed in those studies.
Rather, \cite{2003A&A...398..687B}, for example, used a moments based
method, which does not result in as compact a representation as does the
SVD based method.

In Sect. \ref{sec:back} I start by discussing details of the models used,
the radiative transfer and how the spectral perturbations can be calculated.
In Sect. \ref{sec:fitting} I consider a variety of analysis methods,
starting with the classic cross-correlation. As that is shown to be sub-optimal,
I also discuss the option of performing least squares fits and describe an SVD based method. As these methods result in multiple time series, I finish Sect. \ref{sec:fitting} by outlining
how multiple time series can be analyzed.
Finally I discuss some of the issues still to be addressed and
conclude.

Issues of how the methods may be applied to a given spectrograph,
how they may be made more numerically efficient and the analysis of
simulated or real time series is deferred for later consideration.

\section{Background}
\label{sec:back}

Before addressing the fitting of the spectra I will discuss, in the following
subsections,
the MHD models used, the radiative transfer performed and how those are
used to derive the perturbations to the spectra.

\subsection{Convection models and radiative transfer}
The convection models used here are snapshots from the simulations of
\cite{2013A&A...558A..48B}
with surface properties corresponding to
main sequence stellar models of types F3, G2, K0, K5, M0 and M2
and with sizes ranging from 30~Mm x 30~Mm x 9~Mm for the F3 model to
1.56~Mm x 1.56~Mm x 0.8~Mm for the M2 model.
For each spectral type, models with injected magnetic fields of 0~G, 20~G, 100~G and 500~G were made.
The reader is referred to \cite{2013A&A...558A..48B} for further details of
the various simulations.

For the radiative transfer SPINOR \citep{2000A&A...358.1109F} was used to synthesize various lines for a selection of the models discussed above.

The main line studied is FeI at 6173~\AA\, which is used by
the HMI instrument \citep{2012SoPh..275..229S} and has the advantage that it has simple
atomic properties, a large Land\'e Factor, is relatively free of blends
and that some of the results can be verified by observations.
For selected models the FeI line at 5250~\AA\ and the SiI line at 10827~\AA\ 
were also used.
In addition to these real lines, the 6173~\AA\ line
was also modeled assuming abundances of 0.1 and 10 times solar, to simulate
weaker and stronger lines at otherwise fixed atomic physics.
Finally a line synthesis ("500H") is done for the 500~G case with the magnetic
field artificially set to zero. This makes it possible to disentangle the
effects of the field on the MHD simulations and on the final radiative
transfer calculation.

The line profiles are synthesized for a number of viewing angles
at a spectral resolution of 7.5~m\AA\ with 201 points covering $\pm$0.75~\AA\ 
(except for the SiI 10827~\AA\ line where the spacing is 30~m\AA\ covering $\pm$3~\AA\ in order to accommodate the wider line)
and the resulting profiles are averaged horizontally.

\begin{figure}
\begin{center}
\includegraphics[width=1.00\columnwidth]{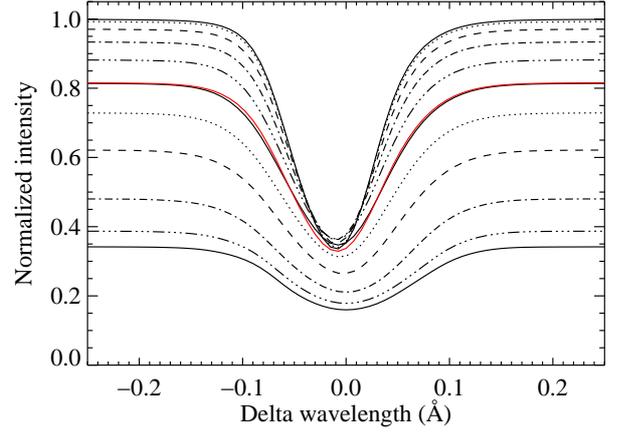}
\end{center}\caption[]{
Line profiles for the FeI 6173~\AA\ line for the G2 star with 0~G. From top to bottom the viewing angles are $0^\circ$, $10^\circ$, ..., $80^\circ$, $85^\circ$ and $87^\circ$.
The profiles were normalized to the disk center continuum.
The red line indicates the disk averaged line profile in the absence of rotation.
}\label{avprofiles}\end{figure}

\begin{figure}
\begin{center}
\includegraphics[width=1.00\columnwidth]{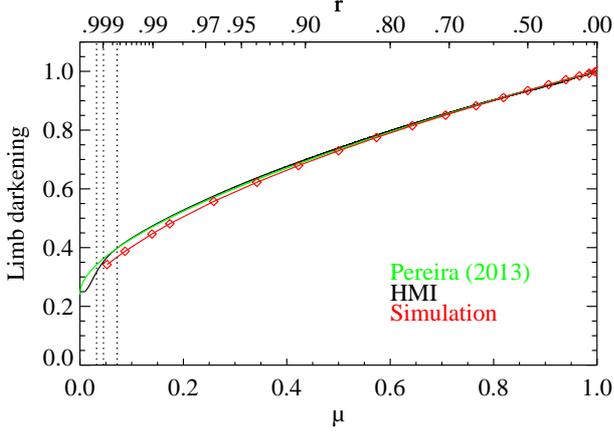}
\end{center}\caption[]{
Continuum limb darkening near the FeI line at 6173~\AA.
Results are shown both as a function of
the fractional solar radius $r$ (on the top axis)
and
$\mu=\sqrt{1-r^2}$ (on the bottom axis).
Red: From the simulations shown in Fig. \ref{avprofiles}, including a number of additional angles.
Black: An estimate from the HMI instrument.
Green: An estimate from \cite{2013A&A...554A.118P} courtesy of R. Trampedach.
Vertical dotted lines indicate 1.0, 2.0 and 5.0 pixels from the limb.
Note that the results were not corrected for the PSF of the HMI instrument.
}\label{limbdark}\end{figure}

Figure \ref{avprofiles} shows the profiles for the G2 star at 0~G.
The limb darkening is clearly visible as a decrease of the continuum
with viewing angle, as is the broadening of the line
towards the limb.
The line also becomes shallower and shifts to the red towards the limb due to convective blueshift.
Of these quantities the limb darkening is the easiest to verify
observationally and
Fig. \ref{limbdark} shows a comparison between the observed limb darkening from HMI and that from various calculations. In general the correspondence is quite good, but there are also differences, especially very close to the limb where the point spread function (PSF) of HMI has not been taken into account.
More importantly, the calculated limb darkening appears to be too steep,
even far from the limb.
Interestingly, the results of \cite{2013A&A...554A.118P}
are significantly closer to the observed results and they
concluded that the key improvement in their calculations is
that the radiative transfer used inside their MHD
simulation was performed more accurately,
resulting in structural changes near the surface.
Such detailed radiative transfer is typically not
performed as part of large MHD simulations due to computational cost,
but was done by \cite{2013A&A...554A.118P} in order to better model
the center-to-limb variations of various observed quantities.
The radiative transfer used for the detailed line synthesis, given
an MHD cube, is believed to be accurate for both their computations
and those performed here.

In any case, the limb darkening is quite well modeled, giving confidence
in the accuracy of the simulations and radiative transfer.

For other stars \cite{2017A&A...605A..91D}, recently demonstrated that the
accuracy of the line profile calculations may be checked by using spectroscopic observations
during planetary transits.

\goodbreak

\subsection{Effects of oscillations on spectra}
To calculate the effect of the oscillations on the observed spectra, one
needs to integrate them over the stellar disk $(\odot)$:
\begin{eqnarray}
I(\lambda,t ) &=& \int_{\rm \odot} I (x,y,\lambda,t) d\vec{r} \nonumber \\
&=& \int_{\rm \odot} I_m (\mu , \lambda+k V_0  +k V^\prime(t)) d\vec{r} \nonumber \\
&\approx& \int_{\rm \odot} I_m (\mu , \lambda+k V_0 ) d\vec{r} + \int_{\rm \odot} k V^\prime(t) I_m^\prime (\mu,\lambda+k V_0) d\vec{r} \nonumber \\
&=& I_0 (\lambda) + k \int_{\rm \odot} V^\prime(t) I_m^\prime (\mu,\lambda+k V_0) d\vec{r} \nonumber \\
&\equiv& I_0 (\lambda) + \delta I(\lambda,t) ,
\label{integrate}
\end{eqnarray}
where $\lambda$ is the wavelength, $t$ time,
$x$ and $y$ are the coordinates relative to disk center (in
units of the stellar radius), $\vec{r} = (x,y)$, $\mu^2=1-x^2-y^2=1-r^2$,
$r$ is the fractional radius,
$I_m$ the model intensity,
$I_m^\prime = dI_m/d\lambda$,
$V_0$ the background stellar LOS velocity
(assumed here to be due to solid body rotation), $V^\prime$ is the LOS velocity induced by the mode
(assumed small), $k=d\lambda/d V = \lambda/c$ and $c$ is the speed of light.
For simplicity the dependencies of $V_0$ and $V^\prime$ on $x$ and $y$
and the fact that $\lambda$ is in reality discretized
are suppressed here and in the rest of this article.
Similarly time variations of $V_0$ have been ignored.
An obvious source for such variations is the Earth's orbital motion which will
cause variations far outside of the linear range and which will have to be
dealt with by shifting the spectra or fitting templates.
It has been assumed that the only change to the spectrum, at a given
spatial location, due
to the mode is that caused by Doppler shift. In other words intensity, linewidth and other thermodynamic
changes are not taken into account.

To calculate the effect of a given mode one needs to calculate the LOS
velocity caused by it.
Assuming that the modes are undistorted
by asphericities (such as rotation or starspots), are undamped and that only the radial component is significant,
one obtains, for a mode with unit amplitude:
\begin{eqnarray}
V^\prime (x,y,t) &=& \Re \left( \mu Y_l^m (x,y) e^{-i\omega_{lm} t}\right) \nonumber \\
&=& \Re \left( \mu P_l^{|m|} (\sin\theta) e^{im\phi-i\omega{_lm} t} \right ) \nonumber\\
&=& \mu P_l^{|m]} (\sin\theta) (\cos\omega_{lm} t \cos m\phi + \sin\omega_{lm} t \sin m\phi) \nonumber \\
&\equiv& V_{lm}^c \cos\omega_{lm} t + V_{lm}^s \sin\omega_{lm} t,
\label{vmode}
\end{eqnarray}
where the factor $\mu$ accounts for the velocity projection factor,
$Y_l^m$ is a spherical harmonic, $P_l^m$ an associated Legendre function,
$x=\sin\phi\cos\theta$, $y=\sin\theta\sin i - \cos\phi\cos\theta\cos i$,
$\phi$ is longitude, $\theta$ is latitude,
$i$ is the inclination of the rotation axis,
and $\omega_{lm}$ is the mode frequency.
With the convention chosen $V_{l-m}^c = V_{lm}^c$ and
$ V_{l-m}^s = -V_{lm}^s$.
Without loss of generality the coordinate system was chosen to have the stellar
rotation axis aligned with the y-axis.

Substituting Eq. (\ref{vmode}) into Eq. (\ref{integrate}) one obtains:
\begin{eqnarray}
\delta I(\lambda,t )
&=& k \int_{\rm \odot} (V_{lm}^c \cos\omega t + V_{lm}^s \sin\omega t) I_m^\prime (\mu,\lambda+k V_0) d\vec{r} \nonumber \\
&\equiv& \delta I_{lm}^c (\lambda)  \cos\omega t + \delta I_{lm}^s (\lambda) \sin\omega t
\label{combo}
\end{eqnarray}
Note that $\delta I_{l-m}^c = \delta I_{lm}^c$ and $\delta I_{l-m}^s = -\delta I_{lm}^s$ and that
in the absence of E-W asymmetries (e.g. rotation) the $\delta I_{lm}^s$ term vanishes.

While I have, for simplicity, assumed that the only background velocity
is that from uniform rotation
and that the modes are undistorted, the formalism can easily accommodate
the more general case. For some details of how this can be done see
\cite{2006A&A...455..227Z}. Similarly it is straightforward to include
thermodynamic changes, when known.

\section{Fitting methods}
\label{sec:fitting}

In this section I will describe the results of three methods for
extracting information from the observed spectra. A simple cross-correlation,
a fit of the expected perturbations and an SVD based analysis.

\subsection{Cross-correlation analysis}
A simple and commonly used method used to extract the Doppler shift
is to cross-correlate
the reference spectrum with the observed one, using methods such as the 
one described by \cite{2001A&A...374..733B}.
In principle the
wavelength dependence of the noise should be taken into account, but
for simplicity I will start by assuming a uniform noise and discuss
the effect of the wavelength dependence of the noise in the next subsection.

As the oscillation
velocities are small and do not cause a significant smearing of the
line (relative to the zero oscillation case) the reference
spectrum can be taken as $I_0$. Given that the perturbations are small, the
cross-correlation is equivalent to a fit of $\delta I$ at each time,
to the derivative $I_0^\prime$ of $I_0$ with respect to $\lambda$,
assuming a uniform error, using the following equation:
\begin{equation}
\delta I(\lambda,t ) = V_{\rm fit}(t) k I_0^\prime (\lambda),
\label{simple-fit}
\end{equation}
where $V_{\rm fit}$ is the fitted velocity. Given Eq. (\ref{combo}), and since a unit amplitude oscillation was assumed, the visibilities, $S$, can be defined
by fitting the perturbations to the derivative:
\begin{equation}
\delta I_{lm}^c (\lambda) = S_{lm}^c k I_0^\prime (\lambda)
\label{simple_sens_c}
\end{equation}
and
\begin{equation}
\delta I_{lm}^s (\lambda) = S_{lm}^s k I_0^\prime (\lambda) ,
\label{simple_sens_s}
\end{equation}
As discussed later, only $S_{lm}^c$ is significant.

\begin{figure}
\begin{center}
\includegraphics[width=1.00\columnwidth]{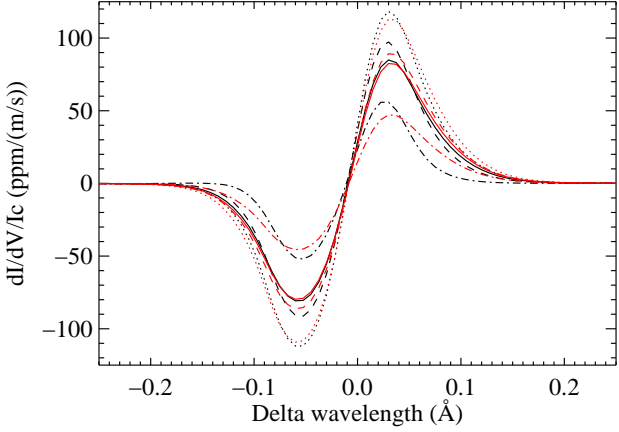}
\end{center}\caption[]{
The perturbations $\delta I_{lm}^c$ (black) for the 6173~\AA\ line for the G2 star
with 0~G,  $i=0^\circ$ and no rotation.
The values of $(l,m)$ are $(0,0)$ (solid),
$(1,0)$ (dotted), $(2,0)$ (dashed) and $(3,0)$ (dash-dotted).
Also shown (in red) is the derivative $I_0^\prime$ scaled to best fit each
perturbation.
The shape (but not magnitude) of the perturbations do not depend on $i$ and $m$ in the absence of rotation.
}\label{dprofiles}\end{figure}

To illustrate how well these profiles match, Fig. \ref{dprofiles}
shows $\delta I_{lm}^c$ for various modes together with fits to $I_0^\prime$.
For the lowest degrees the fit of $I_0^\prime$ is very good but it
starts to disagree more as the degree is increased. 
Still, the approximation used for Eq. (\ref{simple-fit}) is perhaps adequate
to justify using the cross-correlation method for this case.
The decrease of the visibility at higher $l$ is also evident.

\begin{figure}
\begin{center}
\includegraphics[width=1.00\columnwidth]{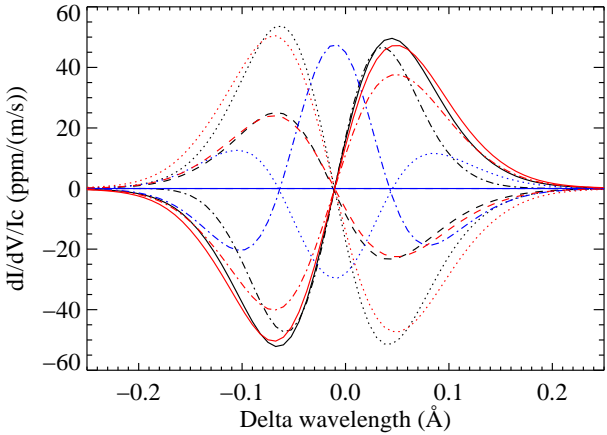}
\end{center}\caption[]{
Similar to Fig. \ref{dprofiles} this shows
the perturbations $\delta I_{lm}^c$ (black) and $\delta I_{lm}^s$ (blue)
for the G2 star with $i=90^\circ$
and solid body rotation with an equatorial rotation velocity of 4~km/s.
The values of $(l,m)$ are $(0,0)$ (solid),
$(1,1)$ (dotted), $(2,0)$ (dashed) and $(2,2)$ (dash-dotted).
Also shown (in red) is the derivative $I_0^\prime$ scaled to best fit each
of the $\delta I_{lm}^c$.
Note that the unperturbed line, and thus the derivative, is different from
the one in Fig. \ref{dprofiles} due to the rotational broadening.
}\label{dprofiles4}\end{figure}

If the star is rotating and not observed from the pole ($i=0^\circ$)
the results change substantially, as illustrated in Fig. \ref{dprofiles4}.
While $\delta I_{lm}^c$ is still moderately well fitted by $I_0^\prime$,
$\delta I_{lm}^s$ is now non-zero and shows a very different profile,
corresponding to a narrowing and widening of the line, which is
not well fitted by the derivative.
I will return to this issue below.

\begin{figure}
\begin{center}
\includegraphics[width=1.00\columnwidth]{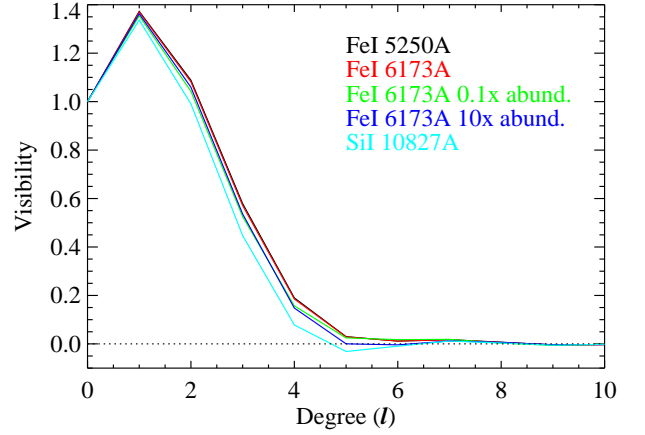}
\end{center}\caption[]{
Visibilities for the $m=0$ modes for the G2 star as a function of
$l$ and spectral line, for the 0~G case and an
inclination angle of $0^\circ$.
In addition to the regular lines, profiles were synthesized using
larger and smaller abundances of Fe for the 6173~\AA\ line (using
the same simulations).
For clarity the visibilities have, for each spectral line, been divided
by the visibility of $l=0$.
Note that the visibilities of the other $m$ values are zero at $i=0^\circ$.
}\label{vis_lines}\end{figure}

\begin{figure}
\begin{center}
\includegraphics[width=1.00\columnwidth]{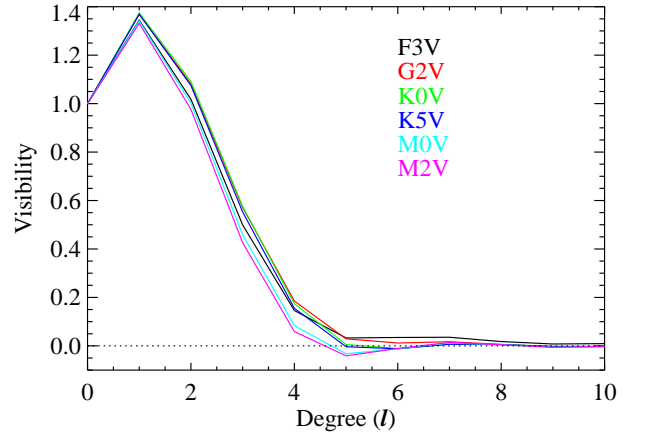}
\end{center}\caption[]{
Similar to Fig. \ref{vis_lines} for different stars.
All cases are for the 6173~\AA\ line without a magnetic field,
no rotation, $i=0^\circ$ and $m=0$
and the visibilities have been divided by the one for $m=l=0$.
}\label{vis_stars}
\end{figure}

Examples of the visibilities $S_{lm}^c$
are shown in Fig. \ref{vis_lines} for a variety
of spectral lines. As can be seen the differences are modest,
especially in the visible, even if the abundances
are changed substantially. The SiI line does show a slightly
different behavior, likely because it is a much broader line and in the near infrared.
As such it should be possible to only model selected lines and interpolate the results.
For different stars (Fig. \ref{vis_stars}) there is a modest trend
with stellar type, but the overall trend remains the same.
Again it would appear that while the variations should not be neglected, it should be possible to interpolate between the spectral types, avoiding the need to run large simulations for each star.

\begin{figure}
\begin{center}
\includegraphics[width=1.00\columnwidth]{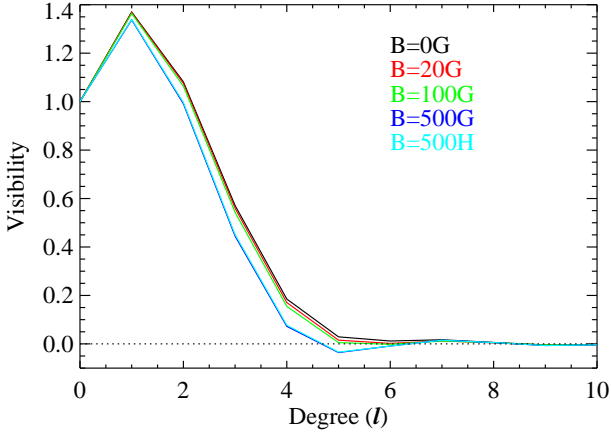}
\end{center}\caption[]{
Similar to Fig. \ref{vis_lines} for the different field strength cases
in \cite{2013A&A...558A..48B}.
The "500H" case used the "500G" simulation but with the field set
to zero in the radiative transfer calculations.
All cases are for the G2 star using the 6173~\AA\ line, no rotation, $i=0^\circ$ and $m=0$
and the visibilities have been divided by the one for $m=l=0$.
}\label{vis_fields}\end{figure}

Adding a modest magnetic field also does not have a dramatic effect
as shown in Fig. \ref{vis_fields}. For the, possibly unrealistic, 500~G
case the change is more significant. It is interesting that
the changes in the structure dominates over the radiative transfer,
as can be seen by turning off the field in the 500~H case, which 
changes the result by much less than did the introduction of the magnetic
field in the MHD simulations.
As such it appears that the presence of magnetic fields can probably be ignored unless the fields are quite strong.

\begin{figure*}
\begin{center}
\includegraphics[width=18cm]{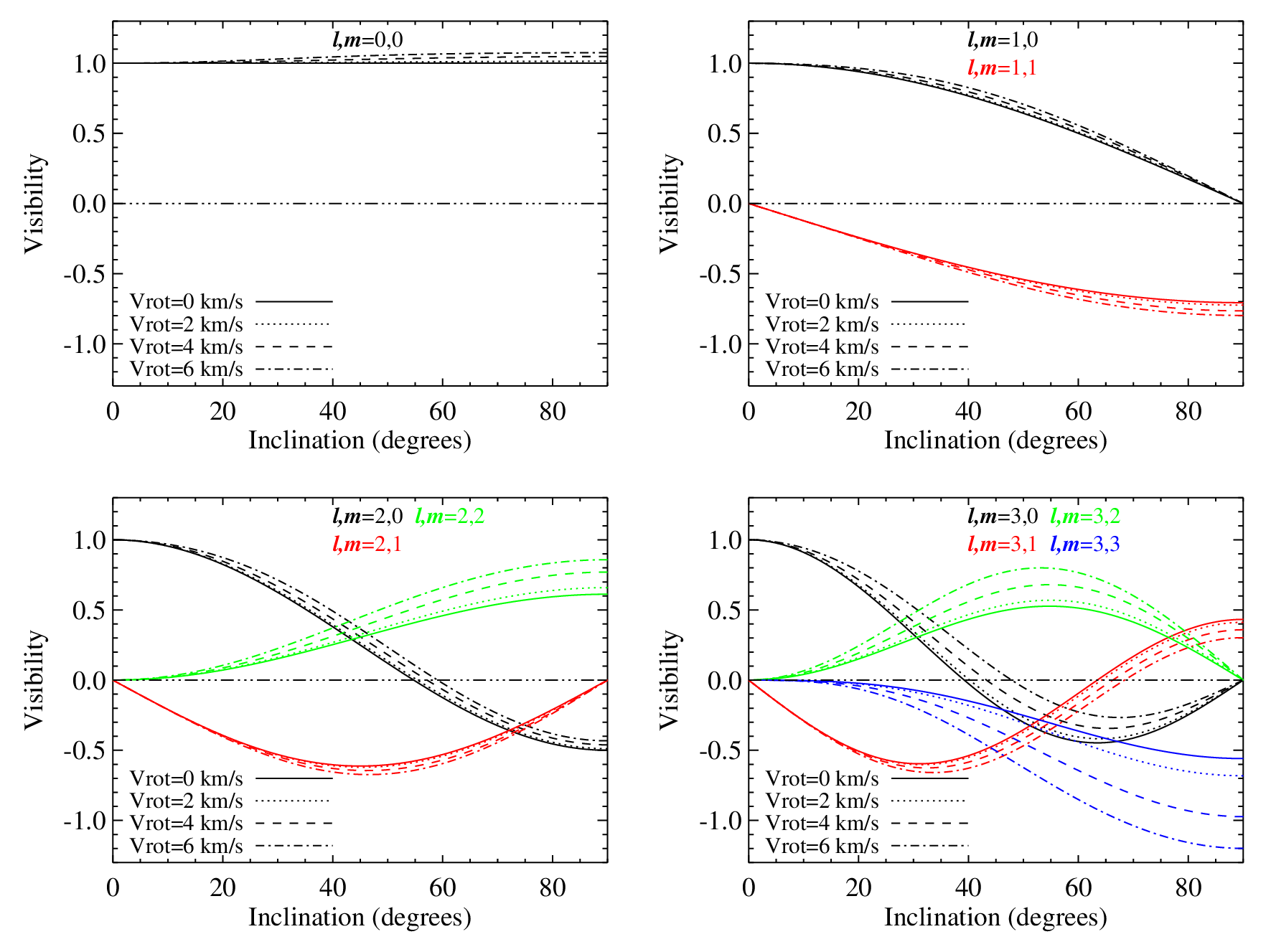}
\end{center}\caption[]{
The visibilities $S_{lm}^c$ of the $l=0$ through $l=3$ modes for the G2 star as a function of
inclination angle and equatorial rotation velocity for the case with 0~G and using the
6173~\AA\ line.
For clarity the values have been divided by the visibility of the 
corresponding $m=0$ mode at $i=0^\circ$.
The corresponding $S_{lm}^s$ values, which are not shown, go up to
0.02, but are generally less than 0.01.
}\label{vis_rot4}\end{figure*}

The inclination dependence of the
visibilities shown in Fig. \ref{vis_rot4} is much more interesting.
Without rotation the inclination dependence follows the
prediction by \cite{2003ApJ...589.1009G} (not shown).
But even for a modest rotation rate of 4~km/s (roughly twice solar)
the deviations are dramatic.
For an equatorial observation ($i=90^\circ$) the
ratio of the visibility of the $(l,m)=(3,3)$ mode relative to that of the
$(3,1)$ mode changes from -1.29 in the absence of rotation to -2.71.

That the rotation has a significant effect on the visibilities
is is not too surprising.
As the rotation increases, the line emitted far from the central meridian moves
far away from that at the center (and the average) and so the 
measurement ceases to be sensitive to the oscillations near the edges.
This also explains why the changes become significant when the
Doppler shift at the east and west limbs becomes comparable to the linewidth
(the FWHM of the average 6173~\AA\ line in the non-rotating case
corresponds to about 5.3~km/s).

\begin{figure}
\begin{center}
\includegraphics[width=0.95\columnwidth]{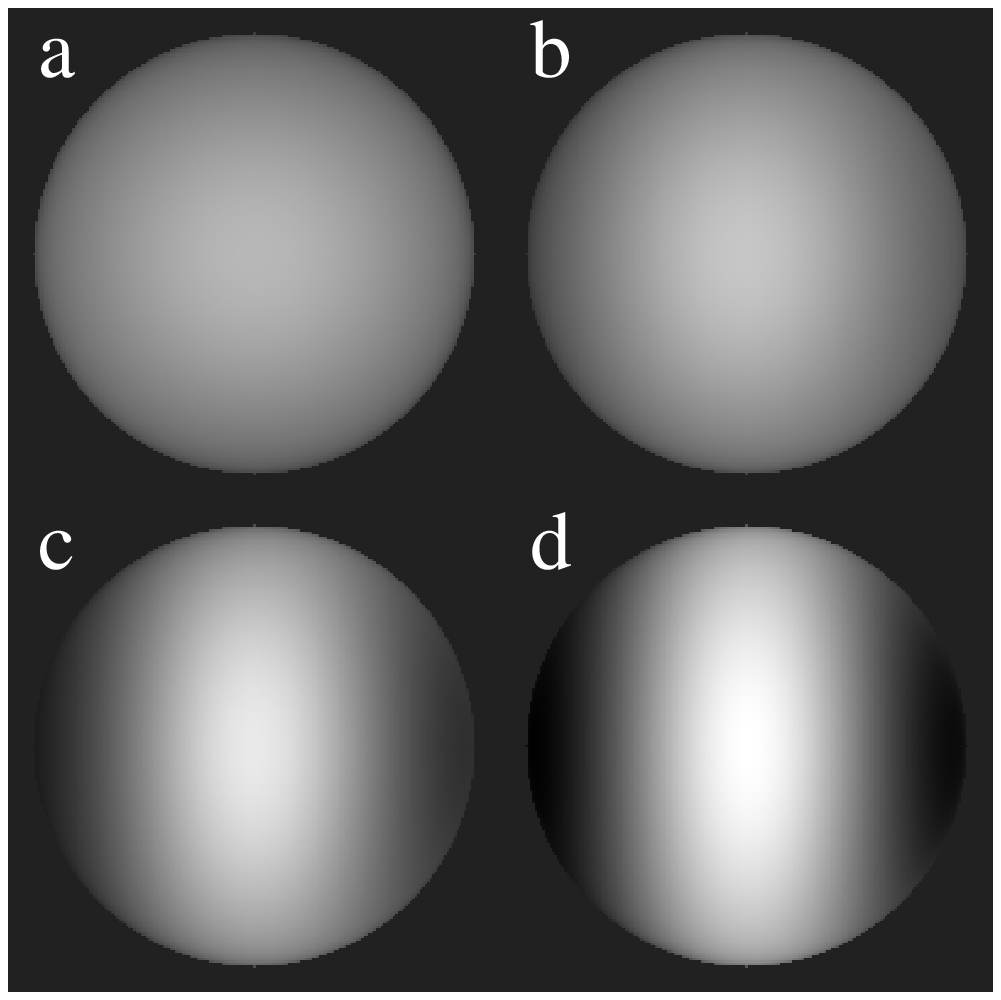}
\end{center}\caption[]{
The sensitivity as a function of position on the disk for
the 6173~\AA\ line for the G2 star
with 0~G and  $i=90^\circ$.
The rotation rates for panels {\bf a} through {\bf d} are 0~km/s, 2~km/s, 4~km/s and 6~km/s.
The maps are normalized to have the same 
integrated sensitivity for the four cases.
Gray scale goes from -0.15 times the maximum sensitivity to the maximum.
}\label{sensmap}\end{figure}

This issue is illustrated in Fig. \ref{sensmap}, where the sensitivity
as a function of position on the stellar disk is shown for an
edge on ($i=90^\circ$) view. As the rotation rate increases the sensitivity
becomes more and more concentrated towards the central meridian.
Beyond about 4~km/s the sensitivity becomes negative at the 
east and west limbs as the shift becomes comparable to the line width.
While barely visible it is also the case that the sensitivity is not
exactly east-west symmetric due to the combination of the rotation and
the convective blueshift.

\begin{figure}
\begin{center}
\includegraphics[width=1.00\columnwidth]{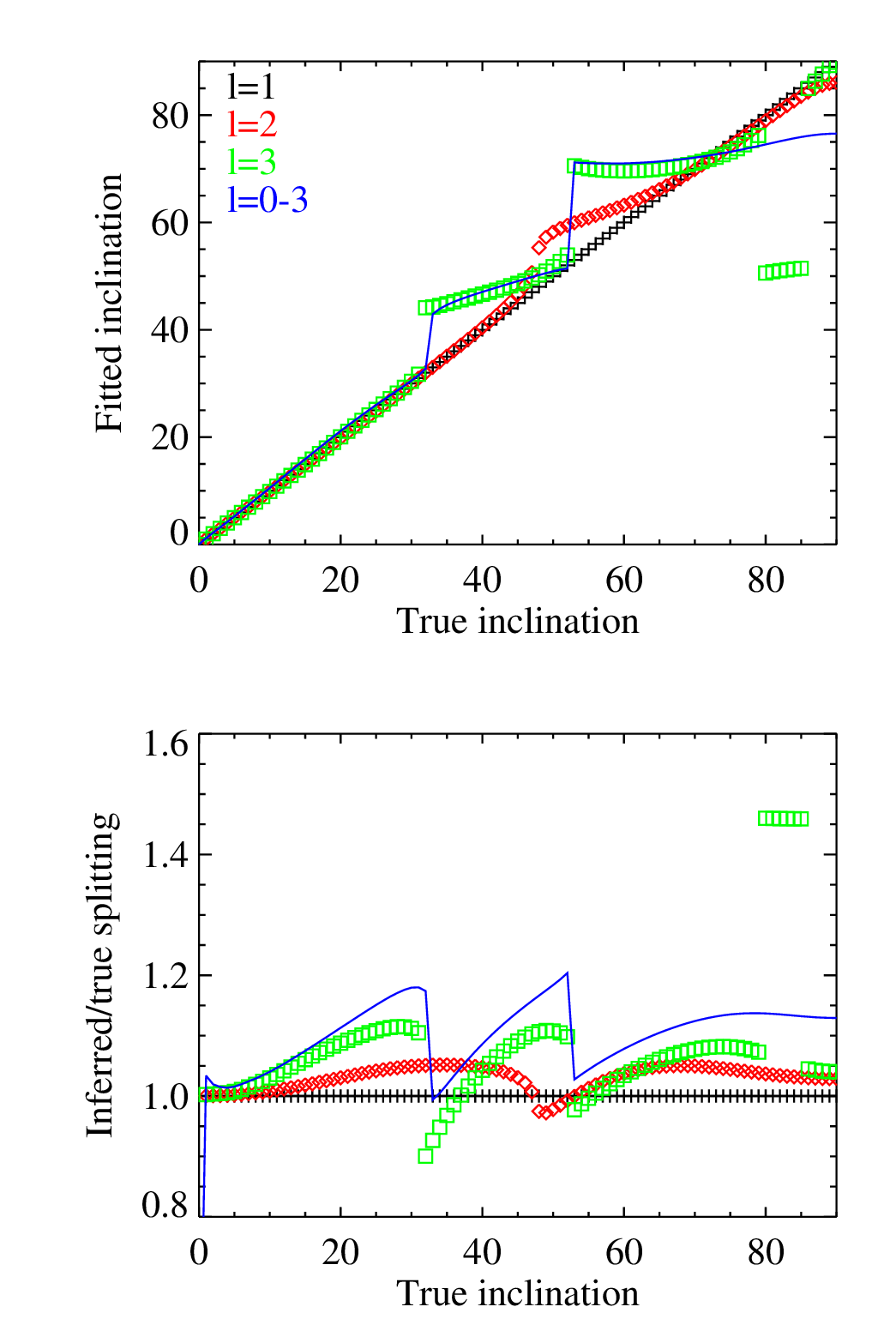}
\end{center}\caption[]{
Results of fitting power spectra constructed with sensitivities calculated including the effects of rotation, but using a model calculated assuming sensitivities ignoring the effects of rotation.
Results are for the G2 star with no field, the 6173 \AA\ line and a rotation velocity of 6~km/s.
The ratio of the FWHM of the modes to the rotation rate is set to 1.0.
The symbols show the results of fitting with one l at a time, the solid line is for using
l from 0 to 3.
In all cases a global search was performed to ensure that the global minimum has been found.
}\label{fit_both}\end{figure}

In solar-like stars the spacing between modes with different $m$ (which is essentially equal to the rotation frequency) is often small or comparable to the linewidth \citep{2007A&A...470..295B,2014A&A...568L..12N}, resulting in the modes blending together.
In this case one therefore has to rely on a model of the visibilities and any errors in that can lead to incorrect mode parameter estimates..
This is illustrated in Fig. \ref{fit_both}.
Here limit spectra calculated using the correct visibilities (those including the effects of rotation) were fitted assuming the incorrect visibilities (those not including the effects of rotation).
As can be seen both the inclinations and the splittings (the spacing between adjacent $m$ values, which is essentially equal to the rotation frequency for Sun-like stars) are mis-estimated, in some cases resulting in values far from the input ones.
Also, the values determined from different values of $l$ are inconsistent,
which may or may not be noticeable.
When the peaks corresponding to the different
values of $m$ are well separated, 
the error will be obvious and would make it clear that there is a problem.

If the correct visibilities are used, the parameters are, of course, correctly estimated.
On the other hand the errors are sub-optimal, as discussed in Sect. \ref{sec:SVD}.

Taking into account the effects of rotation
does not mean that one needs to fit for additional parameters. The
visibilities are given by the rotation rate (and thus the splittings)
and inclination, both of which are already fitted for.
All that needs to be done is to replace the use of the expressions in
\cite{2003ApJ...589.1009G} by a parametrization of the visibilities (e.g. the results shown in Fig. \ref{vis_rot4}), for example by interpolating the values.

\subsection{Least-squares fit}
\label{sec:LS}
As mentioned in the previous subsection, the cross-correlation analysis suffers from several problems.
Among others, the effects of rotation can be quite substantial,
and the implicit assumption that the derivative of the line is a good approximation to the perturbation is thus poor.
Also, it is assumed that the noise is independent of wavelength.
These approximations can be addressed by performing a maximum likelihood fit, which is considered in this subsection.

That the noise is independent of the wavelength is unlikely to be correct,
as discussed in \cite{2001A&A...374..733B}.
A better approximation is to assume that photon noise is the
dominant noise term.
Assuming that the number of photons
is $>>1$ and that the perturbation is small, 
the photon noise can be approximated by a normal distribution
with a variance proportional to the signal.
A fairly benign consequence of this is that an internal
estimate of the noise on the derived velocity, assuming that the
noise equals that in the continuum, will be wrong
relative to that using the photon noise estimate.
In the case of the 6173~\AA\ line by an $l$-independent factor around 1.2.

A more significant problem is that the estimate is statistically
sub optimal, in other words
that an estimate with a better signal to noise ratio can be constructed.
The optimal (maximum likelihood) estimate can be obtained by performing
an error weighted least squares fit, writing the observed spectrum as
\begin{equation}
\label{eq:LS}
I(\lambda,t) = I_0(\lambda) + x_c(t) \delta I_{lm}^c (\lambda) + x_s(t) \delta I_{lm}^s (\lambda),
\end{equation}
thereby obtaining a time series $x_c$ of the cosine component and
another $x_s$ of the sine component.

The error weighted fit being equivalent to an
unweighted fit in which both the data and the model are divided by
the error estimate, which is in this case given by the square root of the model.
A downside of the least squares fitting is that a different function
should ideally be fitted for each mode. This is discussed in the next subsection.

\begin{figure}
\begin{center}
\includegraphics[width=1.00\columnwidth]{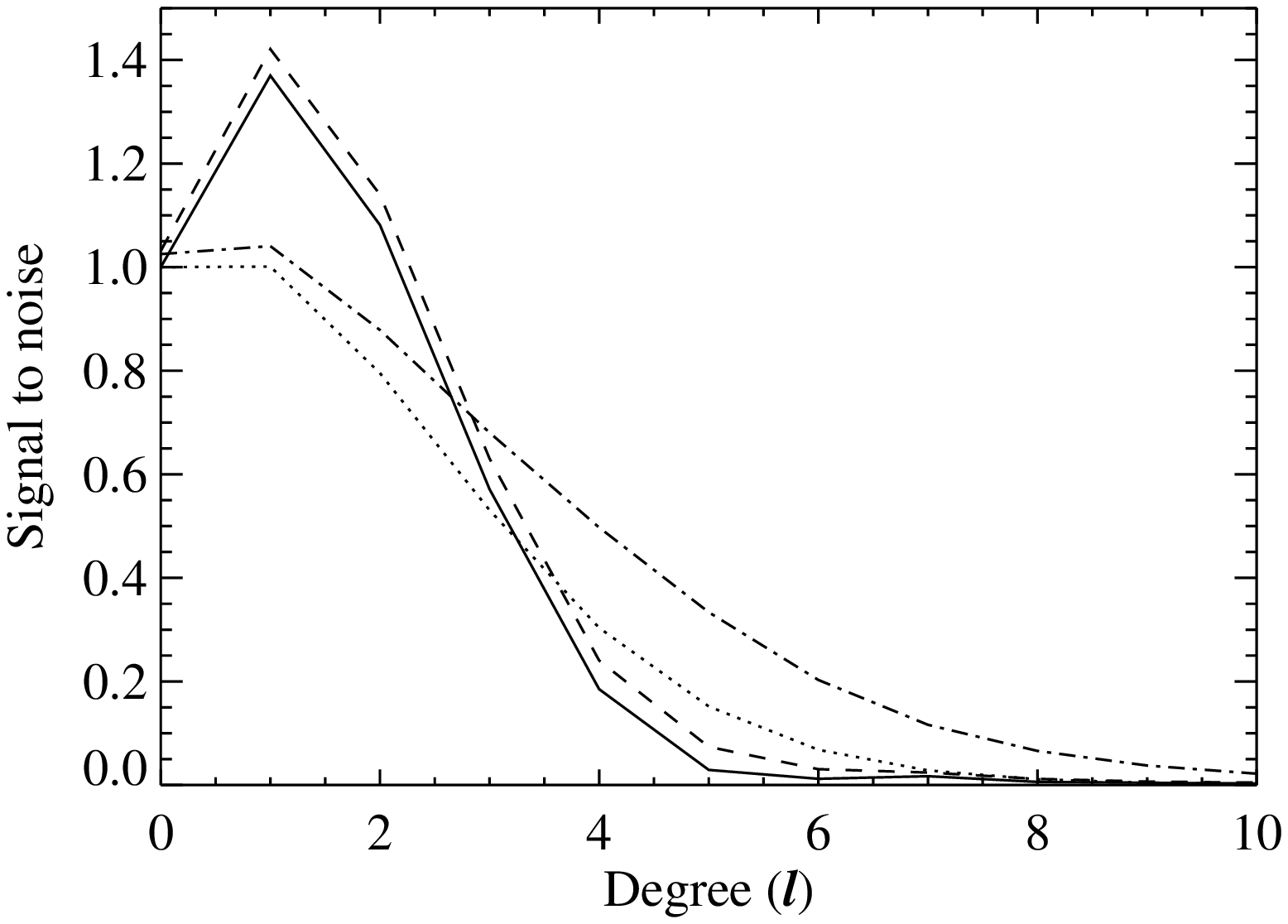}
\end{center}\caption[]{
Signal to noise for various cases. Solid is cross-correlation
is $m=0$ for $i=0^\circ$ and dashed the corresponding fits.
Dotted is the cross-correlation for $m=l$, 4~km/s equatorial rotation
and $i=90^\circ$ while the dash-dotted is the corresponding fit.
In all cases the absolute value of the signal to noise ratio is shown
divided by the corresponding $l=0$ cross-correlation value.
}
\label{vis_opt}
\end{figure}

A comparison between the cross-correlation estimates and the fits 
of only the cosine component (which is dominant at low
rotational velocity) is shown in Fig. \ref{vis_opt}.
Here the signal to noise ratio is shown,
rather than the visibility. As the noise is independent of
the mode in the cross-correlation case, this is proportional to the
absolute value of the visibility.
For the polar ($i=0^\circ$) case, which is unaffected by rotation, the
effect is quite modest. But for an equatorial ($i=90^\circ$) view
the difference is quite significant, especially at higher degree.
At $l=4$ the improvement is roughly a factor of 1.6 and at $l=5$ it is 2.2.

\begin{figure}
\begin{center}
\includegraphics[width=1.00\columnwidth]{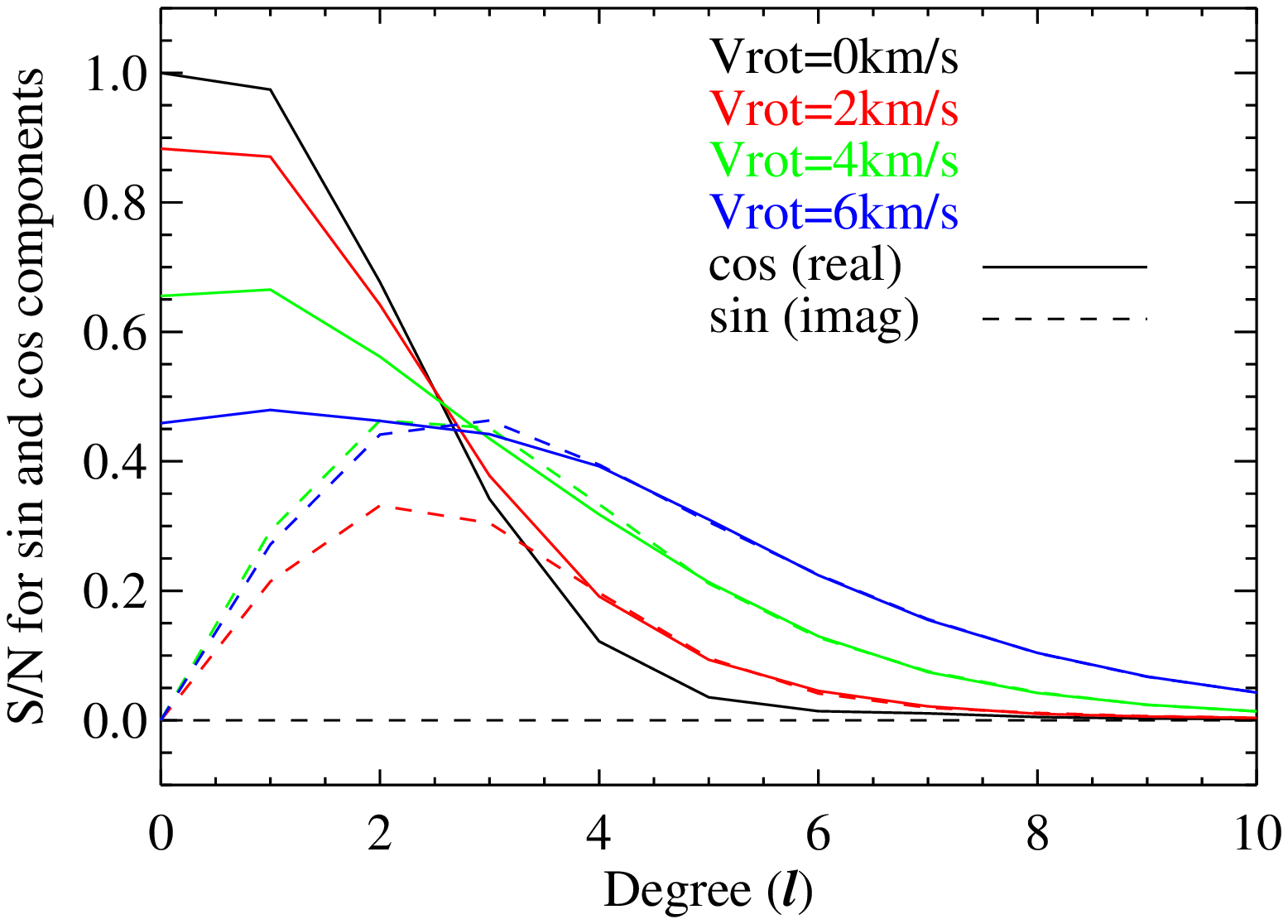}
\end{center}\caption[]{
Signal to noise for the real (cos) and imaginary (sin) components
for $m=l$, $i=90^\circ$ for various equatorial rotation velocities.
The signal to noise ratios were normalized to the $l=0$ case without
rotation.
}
\label{visx_rot}
\end{figure}

Results of fitting for both the sine and cosine components are
shown in Fig. \ref{visx_rot}.
At modest rotation rates the visibility of the real part drops
and the visibility of the imaginary part increases.
At higher degrees the two visibilities become similar.
In other words the broadening and
narrowing of the line caused by the sin component becomes
as significant as the shift of the line caused by the cos component,
which among other things allows for separating prograde and retrograde modes.
In general the ability to observe
two separate time-series opens up the possibility to derive significantly
more information about the modes, but requires a somewhat more complex analysis, as described in Sect. \ref{multi-series}.

While the least squares fitting does provide significant advantages over the cross-correlation
approach, the SVD based approached, discussed in the next subsection, provides significant
advantages, and so the practicalities of implementing the least squares fitting are not
discussed further.

\subsection{SVD analysis}
\label{sec:SVD}
The fact that the perturbations caused by different modes are different
raises a number of questions.
Is it necessary to fit a given spectrum multiple times with functions
designed to fit each mode?
Is it possible to separate more than the cos and sin components, like different $l$ and $m$ values?
A way to address these questions is to perform an SVD (Principal Component)
analysis,
writing the perturbations as:
\begin{equation}
\frac{\delta I_j(\lambda)}{\sqrt{I_0(\lambda)}} = \sum_k U_{k,j} \sigma_k V_k ( \lambda ) ,
\label{SVD}
\end{equation}
where the division by $\sqrt{I_0(\lambda)}$ accounts for the variation
of the photon noise, $j$ encodes the mode ID,
$\sigma_k$ are the singular values,
$U$ the singular vectors in mode ID and
$V$ are the singular vectors in wavelength.
The mode ID could, for example, be $j=(l,m,p)$ where $p$ is
$c$ or $s$ as defined in Eq. (\ref{combo}) at some fixed
inclination and rotation rate.
The properties of the SVD ensure that
the $U$ vectors are orthonormal, as are the $V$ vectors.
By convention, the singular values $\sigma$ are ordered by descending value. Their
squares give the amount of variance of the left hand side of Eq. (\ref{SVD})
that is captured by
the corresponding term (having implicitly assumed that the inherent
mode amplitudes are independent $l$ and $m$, as expected for
solar-like oscillations on a Sun-like star).
An important property of the SVD is that the variance captured by
the first $N$ terms is the maximum possible. No other vectors,
such as the moments used by \cite{2003A&A...398..687B}, can capture more information using $N$ terms.

The vectors $V$ are fitted to the spectra using an unweighted least squares fit of
\begin{equation}
\label{eq:svdfit}
\frac{I(\lambda,t)-I_0(\lambda)}{\sqrt{I_0(\lambda)}} = 
\sum_{k=1}^{k_{max}} y_k(t) V_k(\lambda),
\end{equation}
for some suitable $k_{max}$ (see discussion later in this subsection), resulting in several time series $y_k$.
As the $V$ vectors are orthonormal, the noise on each term is the same and the $y_k$ are given by dot products
\begin{equation}
y_k(t)=\sum_\lambda V_k (\lambda) \frac{I(\lambda,t)-I_0(\lambda)}{\sqrt{I_0(\lambda)}},
\end{equation}
in the case where the fits are done over the same wavelengths as the SVD.

It also follows that the $\sigma U$ are the corresponding visibilities of the various modes.

\begin{figure}
\begin{center}
\includegraphics[width=1.00\columnwidth]{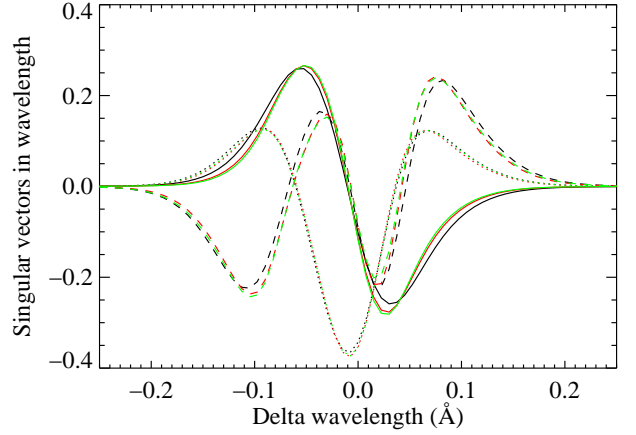}
\end{center}\caption[]{
Singular vectors in wavelength for three cases.
Black: $i=90^\circ$ with 4~km/s equatorial rotation rate.
Red: All inclinations ($i=0^\circ, 10^\circ, ..., 90^\circ$) combined with 4~km/s equatorial rotation rate.
Green: All inclinations combined with all equatorial rotation rates
(0~m/s, 1~km/s, ..., 6~km/s).
Solid, dotted and dashed lines are the first, second and third
vectors, respectively. The signs of some of the vectors were
inverted (they are arbitrary in an SVD).
All cases are for the 6173~\AA\ line at 0~G.
}
\label{svd_plot1}
\end{figure}

Figure \ref{svd_plot1} shows the singular vectors in wavelength for three
different cases and as can be seen the vectors are very similar.
As expected given the earlier results, the
vectors look roughly like the first, second and third derivatives of
the spectral line with wavelength.

\begin{figure}
\begin{center}
\includegraphics[width=1.00\columnwidth]{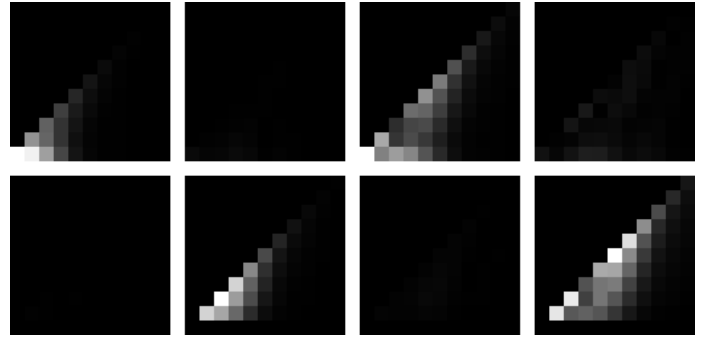}
\end{center}\caption[]{
The first four (from left to right) singular vectors in $l$, $m$ and $p$
for the case show in red in Fig. \ref{svd_plot1}.
Top row is for the cos component and the bottom for the
sin component.
Each block shows the mean over $i$ of the absolute value of the
singular vectors as a function of $l$ (horizontal) and $m$ (vertical)
for $0 \le m \le l \le 10$. The blocks have been individually normalized to have
identical maximum values.
}
\label{svd_plot1b}
\end{figure}

On the other hand the singular values $\sigma$ are somewhat different for the
different cases. A way to quantify this is to
ask how much of the variance (i.e. information in the spectra)
is covered by the lowest few terms.
For an equatorial view and a 4~km/s rotation rate
the numbers are 66.3\%, 94.7\%, 98.7\% and 99.8\%,
for one through four terms.
For a single (4~km/s) rotation rate, but different
viewing angles, the corresponding numbers are 84.6\%, 97.4\%, 99.5\%
and 99.9\%.
For the case with different angles and rotation rates the numbers are
89.3\%, 97.7\%, 99.5\% an 99.8\%.

Given these results it appears that one only needs to fit two or three
terms to the observed spectra in order to extract the vast majority of the
available information.
Exactly how many terms are needed to extract all the statistically
significant information
will depend on the signal to noise ratio, resolution and spectral coverage.
The similarity of the spectra also means that there is little need to
know the inclination and rotation rate in order to fit the spectra, which greatly simplifies the fitting.
How to analyze the multiple resulting time series is discussed in Sect. \ref{multi-series}.

The singular vectors in mode ID ($U$) give the relative
visibility of a given $V$ to various modes, as shown in Fig. \ref{svd_plot1b}.
Given that $V_1$ is very close to the wavelength derivative it is 
not surprising that the corresponding $U$ vector is
dominated by the cos component. Similarly the second vector is dominated
by the sin component and so forth.

\begin{figure}
\begin{center}
\includegraphics[width=1.00\columnwidth]{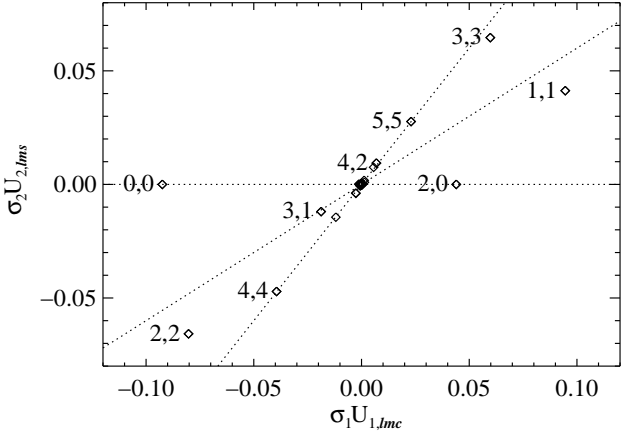}
\end{center}\caption[]{
Visibility of the sin component for the second term in the SVD as
a function of the visibility of the cos component for the first term
for the case shown with black in Fig. \ref{svd_plot1}.
The modes with the largest visibilities are identified by their $l,m$ values.
The steepest dotted line roughly represents the largest ratio, the middle one down half of that and the third one zero.
}
\label{svd_plot2a}
\end{figure}

To further illustrate this Fig. \ref{svd_plot2a}
shows $\sigma_2 U_{2,j}$ for the
sin term as a function of $\sigma_1 U_{1,j}$ for the
cos term.
As can be seen the ratios of the two visibilities
are quite different for different modes, indicating that it may be
possible to partially identify the modes based on this ratio.
Interestingly the ratio depends mostly on the rotation
rate and $(l,m)$ and less on the inclination.

The other two lines studied give similar results. The singular
vectors in wavelength are quite different for the IR line, but
the behavior of the singular vectors in ID and the singular values are similar.
Considering the three lines simultaneously does not lead to a significant
improvement in the ability to separate modes.

It is not practical to perform a brute-force calculation of the perturbations for the entire spectrum. The radiative transfer calculations are extremely
expensive given the size of the computation boxes and the number of lines present in a typical spectrum.
On the other hand the majority of lines behave in a similar way and their
properties vary
slowly with atomic parameters, so it is only necessary to perform a detailed
radiative transfer calculation for a subset. Also, the number of spatial points needed
to adequately sample the convection can likely be reduced dramatically.

Whether it is possible to determine the singular vectors in wavelength
empirically (e.g. by performing an SVD of a number of observed spectra)
is unclear as there are many other sources of variations, such as spectrograph
drift and differential extinction. Also, such an empirical determination
will not directly identify the modes.

Another way to determine the vectors without having to perform the radiative transfer would be to use the method of \cite{2017A&A...605A..91D}, which has the advantage of automatically taking into account the spectrograph PSF, but the obvious disadvantage is that it only works directly for stars with large transiting planets.

Similarly, the fact that the SVD represents the most compact representation
does not mean that this methods has to be used. Other parameterizations, such as
moments \citep{2003A&A...398..687B}, can also be used, but it must be realized that more terms may be
needed and/or that some information is lost.

It is possible that the addition of the thermodynamic perturbations caused by the modes
or some of the other neglected effects will increase the ability to
discriminate between modes, but this will have to await 3D simulations
capable of predicting these effects reliably.

Implementing the fits in practice will require a number of steps. Ideally one would start
by obtaining an (M)HD model for the specific stellar type, synthesize the spectrum as a function
of viewing angle, convolve with the spectrograph PSF, integrating up the perturbations caused by each mode, perform the SVD,
determine how many terms to retain, fitting the spectra using the relevant number of terms
and fitting the resulting time-series, as described in the next subsection.
Unfortunately, many of these steps have complications.

The simulation and spectral synthesis are extremely costly if done brute force.
However, as demonstrated the resulting visibilities only depend weakly on stellar
type and most of the spectral lines of interest are expected to depend only slowly
on the properties of the lines, making it possible to interpolate the results.
Indeed, simply interpolating the results shown in this paper, possibly supplemented with
a few extra stellar models and spectral lines may be adequate, given the limited signal
to noise of the currently available spectra.

The effects of the spectrograph PSF have not been discussed here. If the PSF is well known
it is straightforward to convolve by it after the line synthesis and to check how the
results are affected. If the PSF is poorly known, more research will obviously be needed.

Integrating up the contributions for each mode and performing the SVD should similarly not
present a problem.
Once the S/N for the spectrograph is known, determining the number of terms to retain is
also straightforward to estimate, though one may, of course, choose to try adjacent
values to test the significance.

The fitting of the individual spectra is a linear fit of nearly orthonormal vectors
and should thus be stable and easy to implement.

\subsection{Analysis of multiple simultaneous time-series}
\label{multi-series}

In asteroseismology only a single time series has generally been available for a given star and analysis methods initially developed for Sun as a star observations have been used. For an introduction to the basics, see \cite{1990ApJ...364..699A}.

Clearly, these analysis methods will not be optimal for the multiple simultaneous time series generated using the methods described in Sects. \ref{sec:LS} and \ref{sec:SVD}.
Using the least squares analysis described in Sect. \ref{sec:LS} one obtains two time series ($x_c$ and $x_S$), while  in Sect. \ref{sec:SVD} there can be several time series, depending on the signal to noise ratio.
Thus the question arises of how to go about the case where each mode appears in multiple series, but with different visibilities.
This issue has been addressed extensively in helioseismology and details can be found in e.g. \cite{1992PhDT.......380S} and \cite{2015SoPh..290.3221L}. Here I will briefly outline the general idea and how it might be applied to the present case. I will restrict my case to stochastically driven solar-like oscillations.

In the case of stochastically driven solar-like oscillations, the Fourier transform of a single mode of oscillation has real and
imaginary parts with a mean of zero and a variance given by
\begin{equation}
V(\nu)=\frac{P/w}{1+\left (\frac{\nu-\nu_0}{w} \right )^2},
\end{equation}
where $V$ is the variance, $\nu$ the frequency, $\nu_0$ the mode frequency, $P$ is proportional to the mode power and $w$ is the HWHM of the mode.
Under certain assumptions, most importantly that the mode excitations are frequent and that there are no gaps in the time series, it may be shown that the values of the Fourier transforms are normally distributed and are independent across frequencies and real/imaginary.

When a mode is observed using one of the methods described above, the result is that the resulting Fourier transform are multiplied by complex constants
$C_{jk}$, where $j$ identifies the time-series and $k$ encodes the mode.
The real part of the constant is the sensitivity to the $\cos(m\phi)$ component and the imaginary part the sensitivity to the $\sin(m\phi)$ component, as discussed earlier.
For simplicity, and consistently with the rest of the paper, it is assumed that these constants are independent of frequency and thus the radial order $n$ of the mode.

As we have assumed that the amplitudes are small, it follows that the observed transforms ($\tilde y$) are the sum over the transforms ($\tilde x$) of the individual modes:
\begin{equation}
\tilde y_j (\nu)=\sum_k C_{jk} \tilde x_k(\nu) .
\end{equation}
An important consequence of the fact that a given mode, in general, appears in more than one time-series is that fitting their power spectra (as opposed to Fourier transforms) is sub-optimal. Specifically, the fact that the phases are correlated between the Fourier transforms is ignored resulting in a loss of information when the power spectra are made from the Fourier transforms.

While it is possible to continue using complex numbers (e.g. through the use of cross-spectra), it is perhaps simpler and more intuitive to rearrange the equations to be purely real.
To this end the complex spectra are split into twice as many real spectra to give:
\begin{equation}
\tilde y_j^\prime (\nu)=\sum_k C_{jk}^\prime \tilde x_k^\prime (\nu),
\end{equation}
where $j$ and $k$ now both run over the real and imaginary parts of the observed and mode transforms.

As the $\tilde y_j^\prime$ are sums of normally distributed variables with zero mean, they are themselves normally distributed with zero mean and a covariance matrix with elements given by:
\begin{equation}
E_{nm}({\bf a},\nu_i) = \sum_k C_{nk}^\prime({\bf a})C_{mk}^\prime({\bf a}) V_k ({\bf a},\nu_i) + E_{nm}^{\rm noise} ({\bf a},\nu_i),
\end{equation}
where the vector ${\bf a}$ containing the parameters describing the modes has been made explicit and where a noise term has been added. 
Note the dependence of $E$ on ${\bf a}$ through both $C^\prime$ and $V$.

The probability density for such a multivariate normal distribution is given by:
\begin{equation}
P({\bf a},\nu_i) = |2\pi E({\bf a},\nu_i)|^{-1/2} \exp\left(-\frac{1}{2} y(\nu_i)^T E({\bf a},\nu_i)^{-1}y(\nu_i)\right),
\end{equation}
where $||$ indicates the determinant.

For a maximum likelihood estimate we thus need to minimize minus the logarithm of the product of the probabilities:
\begin{equation}
S({\bf a}) = \sum_i \log | E({\bf a},\nu_i) | + y(\nu_i)^T E({\bf a},\nu_i)^{-1}y(\nu_i),
\end{equation}
where the sum only needs to be made over positive frequencies, given the lack of independent information in the negative part of the transforms.
It is important to note that, despite the superficial similarity, this is not a least squares fit. In a least squares fit, one fits for the mean with a known variance, here the mean is known (zero) and the fit is for the variance.
This results in a non-linear fit, but one that is straightforward to implement from a numerical point of view using standard numerical routines.

Equivalent Bayesian estimates should be straightforward to calculate.

As for implementing such a fit in practice, several things need to be considered. First of all I have not explicitly stated the exact list of parameters ${\bf a}$ used to describe the modes, as a large variety of parameterizations can and have been used, depending on the star and the preferences of various investigators.
Having said that, one would expect that parameterizations similar to those in current use
\citep[e.g.][]{2014A&A...568L..12N}
should work here, which some parameters added to deal with the noise covariance.

Similarly, I have not specified how the noise term could be parameterized, but likely the form of the covariance matrix can be determined from data away from the peaks, as done in helioseismology, given that it would be expected to depend slowly on frequency.

It may be noted that a brute force implementation of the above is actually somewhat inefficient. In particular many of the coefficients in $C^\prime$ may be almost purely real or imaginary and the covariance matrix $E$ will have many identical entries and zeros. However, given that we only have a few time series, the time saved for a maximum likelihood estimate by a more efficient computation is unlikely to be worth the effort. For an MCMC calculation this may not be the case.

\section{Discussion}

The analysis described in this paper is, of course, incomplete and
many effects have been neglected.
For example the stars have been assumed to have no differential rotation
in latitude, the distortion of the eigenfunctions by rotation has been
neglected, the displacement has been assumed to be purely radial and
independent of height, and
it has been assumed that there are no thermodynamic perturbations
accompanying the oscillations.
For a discussion of some of these effects see \cite{2006A&A...455..227Z}.
The assumptions of height independence and lack of thermodynamic perturbations are quite difficult to address numerically,
as the signal to noise ratio of the oscillations in the MHD simulations
is very small due to their limited size and the short simulation time.
For the Sun there is a noticeable height dependent effect \citep{2012ApJ...749L...5Z} in the observations
and \cite{2012ApJ...760L...1B} have shown that the height dependence in the simulations is
far from that predicted by simple theoretical models.
The thermodynamic perturbations are difficult to address
by the MHD models due to the signal to noise limitations, but given the discrepant visibilities reported by
\cite{2014ApJ...782....2L} in intensity and the complexity of the physics it
should not be assumed that the simple analytical models often used give reliable results.
However, for slowly rotating Sun-like stars the effects considered almost
certainly dominate over the ones neglected, at least for spectroscopic
observations.

For photometric observations the situation is different. Here the present
method predicts essentially no perturbations and to obtain reliable
estimates the details of the thermodynamic perturbations with height
have to be understood in detail, as discussed above.

It should also be mentioned that the issues discussed in this paper
are also highly relevant for Doppler imaging and that the two areas
have many similarities.

\section{Conclusion}

In summary it is clear that significant improvements can be made by
careful modeling and analysis of the data and that while some of the
changes are modest they should nonetheless be taken into
account in the analysis of stellar spectra.

Having said that it is also clear that future work is needed
in order to understand some of the remaining uncertainties 
and it would be useful to determine if the observed mode visibilities
agree with those predicted here.

\begin{acknowledgements}
I would like to thank Benjamin Beeck, Regner Trampedach,
Martin Bo Nielsen and Björn Löptien
for useful discussions and help with various calculations.
The HMI data used are courtesy of NASA/SDO and the HMI science team.
\end{acknowledgements}


\bibliographystyle{aa}
\bibliography{vis16}

\begin{thebibliography}{32}
\expandafter\ifx\csname natexlab\endcsname\relax\def\natexlab#1{#1}\fi

\bibitem[{{Anderson} {et~al.}(1990){Anderson}, {Duvall}, \&
  {Jefferies}}]{1990ApJ...364..699A}
{Anderson}, E.~R., {Duvall}, Jr., T.~L., \& {Jefferies}, S.~M. 1990, \apj, 364,
  699

\bibitem[{{Auvergne} {et~al.}(2009){Auvergne}, {Bodin}, {Boisnard}, {Buey},
  {Chaintreuil}, {Epstein}, {Jouret}, {Lam-Trong}, {Levacher}, {Magnan},
  {Perez}, {Plasson}, {Plesseria}, {Peter}, {Steller}, {Tiph{\`e}ne}, {Baglin},
  {Agogu{\'e}}, {Appourchaux}, {Barbet}, {Beaufort}, {Bellenger}, {Berlin},
  {Bernardi}, {Blouin}, {Boumier}, {Bonneau}, {Briet}, {Butler}, {Cautain},
  {Chiavassa}, {Costes}, {Cuvilho}, {Cunha-Parro}, {de Oliveira Fialho},
  {Decaudin}, {Defise}, {Djalal}, {Docclo}, {Drummond}, {Dupuis}, {Exil},
  {Faur{\'e}}, {Gaboriaud}, {Gamet}, {Gavalda}, {Grolleau}, {Gueguen},
  {Guivarc'h}, {Guterman}, {Hasiba}, {Huntzinger}, {Hustaix}, {Imbert},
  {Jeanville}, {Johlander}, {Jorda}, {Journoud}, {Karioty}, {Kerjean},
  {Lafond}, {Lapeyrere}, {Landiech}, {Larqu{\'e}}, {Laudet}, {Le Merrer},
  {Leporati}, {Leruyet}, {Levieuge}, {Llebaria}, {Martin}, {Mazy}, {Mesnager},
  {Michel}, {Moalic}, {Monjoin}, {Naudet}, {Neukirchner}, {Nguyen-Kim},
  {Ollivier}, {Orcesi}, {Ottacher}, {Oulali}, {Parisot}, {Perruchot},
  {Piacentino}, {Pinheiro da Silva}, {Platzer}, {Pontet}, {Pradines},
  {Quentin}, {Rohbeck}, {Rolland}, {Rollenhagen}, {Romagnan}, {Russ}, {Samadi},
  {Schmidt}, {Schwartz}, {Sebbag}, {Smit}, {Sunter}, {Tello}, {Toulouse},
  {Ulmer}, {Vandermarcq}, {Vergnault}, {Wallner}, {Waultier}, \&
  {Zanatta}}]{2009A&A...506..411A}
{Auvergne}, M., {Bodin}, P., {Boisnard}, L., {et~al.} 2009, \aap, 506, 411

\bibitem[{{Baldner} \& {Schou}(2012)}]{2012ApJ...760L...1B}
{Baldner}, C.~S. \& {Schou}, J. 2012, \apjl, 760, L1

\bibitem[{{Bazot} {et~al.}(2007){Bazot}, {Bouchy}, {Kjeldsen}, {Charpinet},
  {Laymand}, \& {Vauclair}}]{2007A&A...470..295B}
{Bazot}, M., {Bouchy}, F., {Kjeldsen}, H., {et~al.} 2007, \aap, 470, 295

\bibitem[{{Bedding}(2014)}]{2014aste.book...60B}
{Bedding}, T.~R. 2014, in Asteroseismology, ed. P.~L. {Pall{\'e}} \&
  C.~{Esteban} (Cambridge, UK: Cambridge University Press), 60

\bibitem[{{Beeck} {et~al.}(2013){Beeck}, {Cameron}, {Reiners}, \&
  {Sch{\"u}ssler}}]{2013A&A...558A..48B}
{Beeck}, B., {Cameron}, R.~H., {Reiners}, A., \& {Sch{\"u}ssler}, M. 2013,
  \aap, 558, A48

\bibitem[{{Borucki} {et~al.}(2010){Borucki}, {Koch}, {Basri}, {Batalha},
  {Brown}, {Caldwell}, {Caldwell}, {Christensen-Dalsgaard}, {Cochran},
  {DeVore}, {Dunham}, {Dupree}, {Gautier}, {Geary}, {Gilliland}, {Gould},
  {Howell}, {Jenkins}, {Kondo}, {Latham}, {Marcy}, {Meibom}, {Kjeldsen},
  {Lissauer}, {Monet}, {Morrison}, {Sasselov}, {Tarter}, {Boss}, {Brownlee},
  {Owen}, {Buzasi}, {Charbonneau}, {Doyle}, {Fortney}, {Ford}, {Holman},
  {Seager}, {Steffen}, {Welsh}, {Rowe}, {Anderson}, {Buchhave}, {Ciardi},
  {Walkowicz}, {Sherry}, {Horch}, {Isaacson}, {Everett}, {Fischer}, {Torres},
  {Johnson}, {Endl}, {MacQueen}, {Bryson}, {Dotson}, {Haas}, {Kolodziejczak},
  {Van Cleve}, {Chandrasekaran}, {Twicken}, {Quintana}, {Clarke}, {Allen},
  {Li}, {Wu}, {Tenenbaum}, {Verner}, {Bruhweiler}, {Barnes}, \&
  {Prsa}}]{2010Sci...327..977B}
{Borucki}, W.~J., {Koch}, D., {Basri}, G., {et~al.} 2010, Science, 327, 977

\bibitem[{{Bouchy} {et~al.}(2001){Bouchy}, {Pepe}, \&
  {Queloz}}]{2001A&A...374..733B}
{Bouchy}, F., {Pepe}, F., \& {Queloz}, D. 2001, \aap, 374, 733

\bibitem[{{Briquet} \& {Aerts}(2003)}]{2003A&A...398..687B}
{Briquet}, M. \& {Aerts}, C. 2003, \aap, 398, 687

\bibitem[{{Brown} {et~al.}(1998){Brown}, {Kotak}, {Horner}, {J.~Kennelly},
  {Korzennik}, {Nisenson}, \& {Noyes}}]{1998ApJS..117..563B}
{Brown}, T.~M., {Kotak}, R., {Horner}, S.~D., {et~al.} 1998, \apjs, 117, 563

\bibitem[{{Chaplin} {et~al.}(2014){Chaplin}, {Basu}, {Huber}, {Serenelli},
  {Casagrande}, {Silva Aguirre}, {Ball}, {Creevey}, {Gizon}, {Handberg},
  {Karoff}, {Lutz}, {Marques}, {Miglio}, {Stello}, {Suran}, {Pricopi},
  {Metcalfe}, {Monteiro}, {Molenda-{\.Z}akowicz}, {Appourchaux},
  {Christensen-Dalsgaard}, {Elsworth}, {Garc{\'{\i}}a}, {Houdek}, {Kjeldsen},
  {Bonanno}, {Campante}, {Corsaro}, {Gaulme}, {Hekker}, {Mathur}, {Mosser},
  {R{\'e}gulo}, \& {Salabert}}]{2014ApJS..210....1C}
{Chaplin}, W.~J., {Basu}, S., {Huber}, D., {et~al.} 2014, \apjs, 210, 1

\bibitem[{{Dravins} {et~al.}(2017){Dravins}, {Ludwig}, {Dahl{\'e}n}, \&
  {Pazira}}]{2017A&A...605A..91D}
{Dravins}, D., {Ludwig}, H.-G., {Dahl{\'e}n}, E., \& {Pazira}, H. 2017, \aap,
  605, A91

\bibitem[{{Fletcher} {et~al.}(2006){Fletcher}, {Chaplin}, {Elsworth}, {Schou},
  \& {Buzasi}}]{2006MNRAS.371..935F}
{Fletcher}, S.~T., {Chaplin}, W.~J., {Elsworth}, Y., {Schou}, J., \& {Buzasi},
  D. 2006, \mnras, 371, 935

\bibitem[{{Frutiger} {et~al.}(2000){Frutiger}, {Solanki}, {Fligge}, \&
  {Bruls}}]{2000A&A...358.1109F}
{Frutiger}, C., {Solanki}, S.~K., {Fligge}, M., \& {Bruls}, J.~H.~M.~J. 2000,
  \aap, 358, 1109

\bibitem[{{Gizon} \& {Solanki}(2003)}]{2003ApJ...589.1009G}
{Gizon}, L. \& {Solanki}, S.~K. 2003, \apj, 589, 1009

\bibitem[{{Grundahl} {et~al.}(2009){Grundahl}, {Christensen-Dalsgaard},
  {Arentoft}, {Frandsen}, {Kjeldsen}, {J{\o}rgensen}, \&
  {Kj{\ae}rgaard}}]{2009CoAst.158..345G}
{Grundahl}, F., {Christensen-Dalsgaard}, J., {Arentoft}, T., {et~al.} 2009,
  Communications in Asteroseismology, 158, 345

\bibitem[{{Grundahl} {et~al.}(2017){Grundahl}, {Fredslund Andersen},
  {Christensen-Dalsgaard}, {Antoci}, {Kjeldsen}, {Handberg}, {Houdek},
  {Bedding}, {Pall{\'e}}, {Jessen-Hansen}, {Silva Aguirre}, {White},
  {Frandsen}, {Albrecht}, {Andersen}, {Arentoft}, {Brogaard}, {Chaplin},
  {Harps{\o}e}, {J{\o}rgensen}, {Karovicova}, {Karoff}, {Kj{\ae}rgaard
  Rasmussen}, {Lund}, {Sloth Lundkvist}, {Skottfelt}, {Norup S{\o}rensen},
  {Tronsgaard}, \& {Weiss}}]{2017ApJ...836..142G}
{Grundahl}, F., {Fredslund Andersen}, M., {Christensen-Dalsgaard}, J., {et~al.}
  2017, \apj, 836, 142

\bibitem[{{Larson} \& {Schou}(2015)}]{2015SoPh..290.3221L}
{Larson}, T.~P. \& {Schou}, J. 2015, \solphys, 290, 3221

\bibitem[{{Lund} {et~al.}(2014){Lund}, {Kjeldsen}, {Christensen-Dalsgaard},
  {Handberg}, \& {Silva Aguirre}}]{2014ApJ...782....2L}
{Lund}, M.~N., {Kjeldsen}, H., {Christensen-Dalsgaard}, J., {Handberg}, R., \&
  {Silva Aguirre}, V. 2014, \apj, 782, 2

\bibitem[{{Lund} {et~al.}(2017{\natexlab{a}}){Lund}, {Silva Aguirre}, {Davies},
  {Chaplin}, {Christensen-Dalsgaard}, {Houdek}, {White}, {Bedding}, {Ball},
  {Huber}, {Antia}, {Lebreton}, {Latham}, {Handberg}, {Verma}, {Basu},
  {Casagrande}, {Justesen}, {Kjeldsen}, \& {Mosumgaard}}]{2017ApJ...835..172L}
{Lund}, M.~N., {Silva Aguirre}, V., {Davies}, G.~R., {et~al.}
  2017{\natexlab{a}}, \apj, 835, 172

\bibitem[{{Lund} {et~al.}(2017{\natexlab{b}}){Lund}, {Silva Aguirre}, {Davies},
  {Chaplin}, {Christensen-Dalsgaard}, {Houdek}, {White}, {Bedding}, {Ball},
  {Huber}, {Antia}, {Lebreton}, {Latham}, {Handberg}, {Verma}, {Basu},
  {Casagrande}, {Justesen}, {Kjeldsen}, \& {Mosumgaard}}]{2017ApJ...850..110L}
{Lund}, M.~N., {Silva Aguirre}, V., {Davies}, G.~R., {et~al.}
  2017{\natexlab{b}}, \apj, 850, 110

\bibitem[{{Nielsen} {et~al.}(2014){Nielsen}, {Gizon}, {Schunker}, \&
  {Schou}}]{2014A&A...568L..12N}
{Nielsen}, M.~B., {Gizon}, L., {Schunker}, H., \& {Schou}, J. 2014, \aap, 568,
  L12

\bibitem[{{Pereira} {et~al.}(2013){Pereira}, {Asplund}, {Collet}, {Thaler},
  {Trampedach}, \& {Leenaarts}}]{2013A&A...554A.118P}
{Pereira}, T.~M.~D., {Asplund}, M., {Collet}, R., {et~al.} 2013, \aap, 554,
  A118

\bibitem[{{Rauer} {et~al.}(2014){Rauer}, {Catala}, {Aerts}, {Appourchaux},
  {Benz}, {Brandeker}, {Christensen-Dalsgaard}, {Deleuil}, {Gizon}, {Goupil},
  {G{\"u}del}, {Janot-Pacheco}, {Mas-Hesse}, {Pagano}, {Piotto}, {Pollacco},
  {Santos}, {Smith}, {Su{\'a}rez}, {Szab{\'o}}, {Udry}, {Adibekyan}, {Alibert},
  {Almenara}, {Amaro-Seoane}, {Eiff}, {Asplund}, {Antonello}, {Barnes},
  {Baudin}, {Belkacem}, {Bergemann}, {Bihain}, {Birch}, {Bonfils}, {Boisse},
  {Bonomo}, {Borsa}, {Brand{\~a}o}, {Brocato}, {Brun}, {Burleigh}, {Burston},
  {Cabrera}, {Cassisi}, {Chaplin}, {Charpinet}, {Chiappini}, {Church},
  {Csizmadia}, {Cunha}, {Damasso}, {Davies}, {Deeg}, {D{\'{\i}}az}, {Dreizler},
  {Dreyer}, {Eggenberger}, {Ehrenreich}, {Eigm{\"u}ller}, {Erikson}, {Farmer},
  {Feltzing}, {de Oliveira Fialho}, {Figueira}, {Forveille}, {Fridlund},
  {Garc{\'{\i}}a}, {Giommi}, {Giuffrida}, {Godolt}, {Gomes da Silva},
  {Granzer}, {Grenfell}, {Grotsch-Noels}, {G{\"u}nther}, {Haswell}, {Hatzes},
  {H{\'e}brard}, {Hekker}, {Helled}, {Heng}, {Jenkins}, {Johansen},
  {Khodachenko}, {Kislyakova}, {Kley}, {Kolb}, {Krivova}, {Kupka}, {Lammer},
  {Lanza}, {Lebreton}, {Magrin}, {Marcos-Arenal}, {Marrese}, {Marques},
  {Martins}, {Mathis}, {Mathur}, {Messina}, {Miglio}, {Montalban}, {Montalto},
  {Monteiro}, {Moradi}, {Moravveji}, {Mordasini}, {Morel}, {Mortier},
  {Nascimbeni}, {Nelson}, {Nielsen}, {Noack}, {Norton}, {Ofir}, {Oshagh},
  {Ouazzani}, {P{\'a}pics}, {Parro}, {Petit}, {Plez}, {Poretti}, {Quirrenbach},
  {Ragazzoni}, {Raimondo}, {Rainer}, {Reese}, {Redmer}, {Reffert},
  {Rojas-Ayala}, {Roxburgh}, {Salmon}, {Santerne}, {Schneider}, {Schou},
  {Schuh}, {Schunker}, {Silva-Valio}, {Silvotti}, {Skillen}, {Snellen}, {Sohl},
  {Sousa}, {Sozzetti}, {Stello}, {Strassmeier}, {{\v S}vanda}, {Szab{\'o}},
  {Tkachenko}, {Valencia}, {Van Grootel}, {Vauclair}, {Ventura}, {Wagner},
  {Walton}, {Weingrill}, {Werner}, {Wheatley}, \&
  {Zwintz}}]{2014ExA....38..249R}
{Rauer}, H., {Catala}, C., {Aerts}, C., {et~al.} 2014, Experimental Astronomy,
  38, 249

\bibitem[{{Ricker} {et~al.}(2014){Ricker}, {Winn}, {Vanderspek}, {Latham},
  {Bakos}, {Bean}, {Berta-Thompson}, {Brown}, {Buchhave}, {Butler}, {Butler},
  {Chaplin}, {Charbonneau}, {Christensen-Dalsgaard}, {Clampin}, {Deming},
  {Doty}, {De Lee}, {Dressing}, {Dunham}, {Endl}, {Fressin}, {Ge}, {Henning},
  {Holman}, {Howard}, {Ida}, {Jenkins}, {Jernigan}, {Johnson}, {Kaltenegger},
  {Kawai}, {Kjeldsen}, {Laughlin}, {Levine}, {Lin}, {Lissauer}, {MacQueen},
  {Marcy}, {McCullough}, {Morton}, {Narita}, {Paegert}, {Palle}, {Pepe},
  {Pepper}, {Quirrenbach}, {Rinehart}, {Sasselov}, {Sato}, {Seager},
  {Sozzetti}, {Stassun}, {Sullivan}, {Szentgyorgyi}, {Torres}, {Udry}, \&
  {Villasenor}}]{2014SPIE.9143E..20R}
{Ricker}, G.~R., {Winn}, J.~N., {Vanderspek}, R., {et~al.} 2014, in \procspie,
  Vol. 9143, Space Telescopes and Instrumentation 2014: Optical, Infrared, and
  Millimeter Wave, 914320

\bibitem[{{Schou}(1992)}]{1992PhDT.......380S}
{Schou}, J. 1992, PhD thesis, Aarhus University, Aarhus, Denmark

\bibitem[{{Schou}(2014)}]{2014IAUS..301..481S}
{Schou}, J. 2014, in IAU Symposium, Vol. 301, Precision Asteroseismology, ed.
  J.~A. {Guzik}, W.~J. {Chaplin}, G.~{Handler}, \& A.~{Pigulski}, 481--482

\bibitem[{{Schou} {et~al.}(2012){Schou}, {Scherrer}, {Bush}, {Wachter},
  {Couvidat}, {Rabello-Soares}, {Bogart}, {Hoeksema}, {Liu}, {Duvall}, {Akin},
  {Allard}, {Miles}, {Rairden}, {Shine}, {Tarbell}, {Title}, {Wolfson},
  {Elmore}, {Norton}, \& {Tomczyk}}]{2012SoPh..275..229S}
{Schou}, J., {Scherrer}, P.~H., {Bush}, R.~I., {et~al.} 2012, \solphys, 275,
  229

\bibitem[{{Stello} {et~al.}(2013){Stello}, {Huber}, {Bedding}, {Benomar},
  {Bildsten}, {Elsworth}, {Gilliland}, {Mosser}, {Paxton}, \&
  {White}}]{2013ApJ...765L..41S}
{Stello}, D., {Huber}, D., {Bedding}, T.~R., {et~al.} 2013, \apjl, 765, L41

\bibitem[{{Walker} {et~al.}(2003){Walker}, {Matthews}, {Kuschnig}, {Johnson},
  {Rucinski}, {Pazder}, {Burley}, {Walker}, {Skaret}, {Zee}, {Grocott},
  {Carroll}, {Sinclair}, {Sturgeon}, \& {Harron}}]{2003PASP..115.1023W}
{Walker}, G., {Matthews}, J., {Kuschnig}, R., {et~al.} 2003, \pasp, 115, 1023

\bibitem[{{Zhao} {et~al.}(2012){Zhao}, {Nagashima}, {Bogart}, {Kosovichev}, \&
  {Duvall}}]{2012ApJ...749L...5Z}
{Zhao}, J., {Nagashima}, K., {Bogart}, R.~S., {Kosovichev}, A.~G., \& {Duvall},
  Jr., T.~L. 2012, \apjl, 749, L5

\bibitem[{{Zima}(2006)}]{2006A&A...455..227Z}
{Zima}, W. 2006, \aap, 455, 227

\end{thebibliography}

\end{document}